\documentclass[final,3p,times]{elsarticle}

\usepackage{amssymb}
\usepackage{amsmath}
\usepackage{soul} 

\usepackage[
    colorlinks=true,
    linkcolor=blue,
    citecolor=blue,
    urlcolor=blue
]{hyperref}

\usepackage{doi}

\journal{Journal of Subatomic Particles and Cosmology}

\begin{document}

\begin{frontmatter}



\title{An Extended Chiral Mean-Field Model with Medium-Modified Thermal Mesons for Hot and Dense Hadronic Matter}

\author[aaa]{Micheal Kahangirwe}\ead{mkahangi@kent.edu}
\author[aaa,ddd]{Rajesh Kumar}\ead{rajesh.sism@gmail.com}
\author[aaa,bbb]{Joaquin Grefa}
\author[bbb]{Konstantin Maslov}
\author[aaa]{Yuhan Wang}
\author[nnn]{Arvind Kumar}
\author[bbb]{Claudia Ratti }
\author[aaa]{Veronica Dexheimer }
\affiliation[aaa]{organization={Center for Nuclear Research, Department of Physics, Kent State University},
             city={Kent},
             postcode={44242},
             state={OH},
             country={USA}}
 \affiliation[bbb]{organization={Department of Physics, University of Houston},
             city={Houston},
             postcode={77204},
             state={TX},
             country={USA}}
 \affiliation[ddd]{Department of Physics, MRPD Government College Talwara,
             city={Punjab},
             postcode={144216},
             country={India}}
\affiliation[nnn]{Department of Physics, Dr. B.R. Ambedkar National Institute of Technology Jalandhar,
             city={Punjab},
             postcode={144011},
             country={India}}

\begin{abstract}
In this conference proceeding we review an extension of the Chiral Mean Field (CMF) model that consistently incorporates interacting thermal mesons with self-consistent in-medium masses. In this approach, the in-medium masses of pseudoscalar and vector mesons are evaluated through the explicit chiral symmetry-breaking and
vector-interaction terms in the Lagrangian respectively, before applying the mean-field approximation. These medium-modified meson properties
introduce additional feedback into the CMF equations of motion, leading
to a revised equation of state. The impact of this refinement is analyzed by comparing the hadronic model
predictions with recent lattice QCD data and other hadronic descriptions, such as the hadron resonance gas model. The modified CMF framework, featuring an improved meson treatment (mCMF),
demonstrates enhanced consistency with lattice-QCD results for thermodynamic observables across
a broad range of temperatures and baryon chemical potentials. 
\end{abstract}




\end{frontmatter}



\section*{Introduction}

Understanding the thermodynamic properties of strongly interacting matter remains one of the major goals of nuclear physics. While lattice QCD provides reliable predictions for QCD thermodynamics at vanishing baryon chemical potential, the fermion sign problem limits direct calculations at finite density \cite{Ratti:2018ksb}. Current lattice-QCD extrapolation schemes are limited to approximately $\mu_B/T\lesssim3.5$ \cite{Abuali:2025tbd}. Consequently, effective models remain indispensable for studying dense matter encountered in heavy-ion collisions, neutron stars, and binary neutron-star mergers.
Among these approaches, the Chiral Mean Field (CMF) model successfully reproduces nuclear saturation properties, finite nuclei, neutron-star observations, and heavy-ion phenomenology while providing a thermodynamically consistent equation of state over a broad range of densities and temperatures \cite{Cruz-Camacho:2024odu}. The model incorporates spontaneous and explicit chiral symmetry breaking  while baryons interact via scalar and vector meson mean fields.
Despite these successes, previous implementations of the CMF model treated thermal mesons as non-interacting particles with fixed vacuum masses. Near the chiral crossover, thermal mesons become increasingly abundant and contribute significantly  to the hadronic medium \cite{Kumar:2025rxj}. In-medium modifications of meson masses are expected from partial restoration of chiral symmetry \cite{Jido:PhysRevC.85.032201} and should influence the thermodynamics.
The recently developed Chiral Mean Field  model with improved meson description (mCMF) addresses this limitation by introducing interacting thermal mesons whose masses depend self-consistently on the surrounding medium. Unlike previous hybrid approaches, the new formulation computes meson masses directly from the effective Lagrangian of the nonlinear SU(3) sigma model before applying the mean-field approximation, allowing thermal mesons to modify the equation of state (EoS) through feedback contributions in the equations of motion. In this work, we discuss how these medium-modified thermal mesons affect hadronic thermodynamics and compare the resulting equation of state with lattice-QCD results.

\section*{ Chiral Mean Field with improved meson  Formalism}
\label{subsec1}
The Chiral Mean Field (CMF) model is formulated from the nonlinear realization of the SU(3) sigma model that incorporates the spontaneous and explicit chiral symmetry breaking, and broken scale invariance of Quantum Chromodynamics (QCD). It can be written as
\begin{equation}
\mathcal{L}_{\mathrm{CMF}}= \mathcal{L}_{\mathrm{kin}} + \mathcal{L}_{\mathrm{int}} + \mathcal{L}_{\mathrm{scal}} + \mathcal{L}_{\mathrm{vec}} + \mathcal{L}_{\mathrm{SB}} + \mathcal{L}_{\Phi},
\end{equation}
where $\mathcal{L}_{\mathrm{kin}}$ and $\mathcal{L}_{\mathrm{int}}$ denote the kinetic term and describe the interactions between baryons and mesons respectively, $\mathcal{L}_{\mathrm{scal}}$ and $\mathcal{L}_{\mathrm{vec}}$ represent the scalar and vector meson self-interaction terms, $\mathcal{L}_{\mathrm{SB}}$ accounts for explicit chiral symmetry breaking, and $\mathcal{L}_{\Phi}$ is the Polyakov-loop like potential responsible for modeling the confinement--deconfinement transition. 

The thermodynamics of strongly interacting matter is obtained by solving the previous Lagrangian  within the mean-field approximation (MFA). In this approach, the system is assumed to be homogeneous, isotropic, and parity conserving, allowing the meson fields to be replaced by their time and space independent expectation values (summarized in Eq.~\eqref{eq.MFA}). Consequently, only the scalar mesons and the time-like component of the vector mesons are non-zero, while pseudoscalar, axial-vector, and space-like vector fields vanish. This approximation significantly simplifies the many-body problem while preserving the essential physics of chiral symmetry and dense nuclear matter, 

\begin{equation}
    \sigma \longrightarrow \sigma_0, ~~~~V^\mu \longrightarrow \left(V_0,0\right), ~~~~~ <\pi_i>=0,
    \label{eq.MFA}
\end{equation}
In the present work \cite{Kumar:2025rxj}, we restrict ourselves to the hadronic sector (where quark degrees of freedom are neglected) and review an extension of the field-redefined CMF model in Ref. \cite{Kumar:2024owe} by including a thermal meson contribution that includes interactions, meaning that the pseudoscalar and vector meson masses are now dependent on the in-medium meson mean fields.

\begin{figure}[!h]
    \centering
    \includegraphics[width=1\linewidth,
        trim={0cm 0cm 0cm 0.2cm},
        clip]{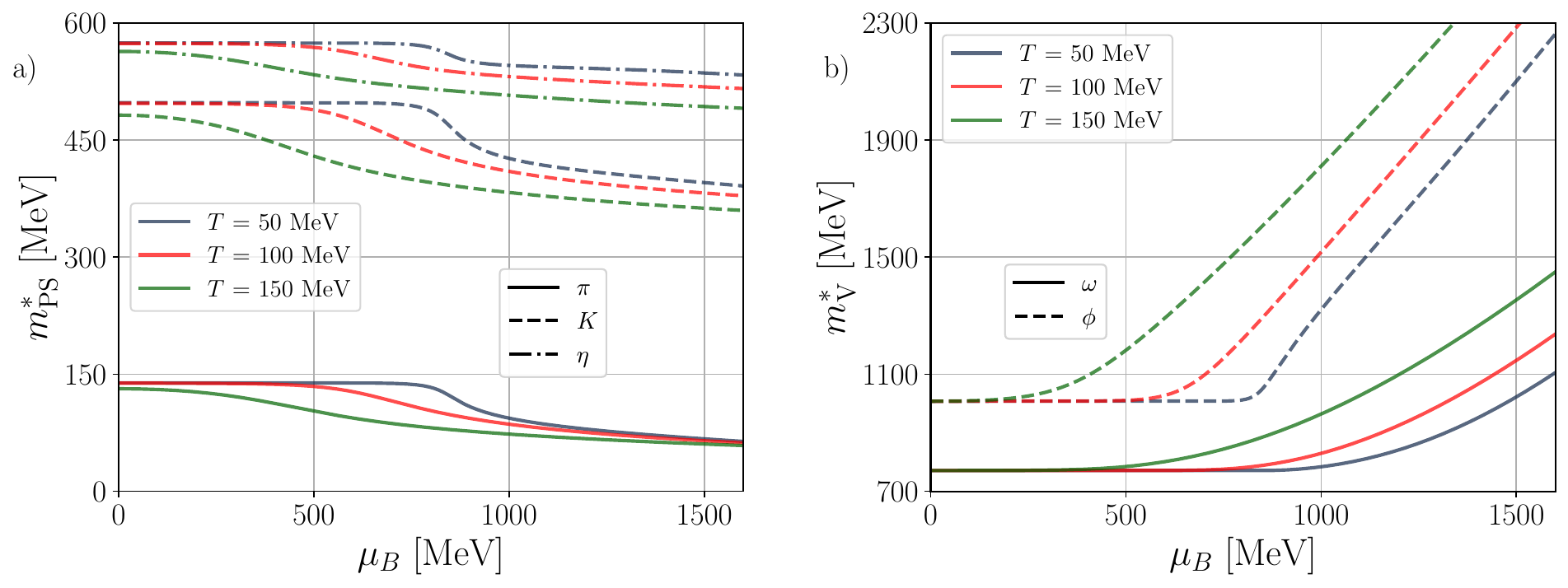}
    \caption{In-medium masses of mesons vs. baryon chemical potential for the interacting meson (mCMF) case. Panels a) and  b), show pseudoscalar and vector mesons respectively, and compare for different values of temperature, $T=50, 100$ and $150 $ MeV.
    (Figure taken from Ref. \cite{Kumar:2025rxj})}
    \label{fig:masses}
\end{figure}

The resulting hadronic (H) grand canonical potential density can be written as:
$$\Omega_H = U +  \Omega_B^{th} + \Omega_M^{th},$$ 
where $U$ contains the meson mean-field interactions, while $\Omega_B^{th}$  and $\Omega_M^{th}$  are thermal contributions for baryons and mesons respectively. 
The thermal mesons now contribute to the minimization of the thermodynamic potential through the dependence of their effective masses on the meson mean fields, including a thermal meson contribution to the minimization conditions of the thermodynamic potential and, therefore, affecting the self-consistent solution for the mean fields.

\begin{equation}
    \frac{\partial(\Omega_H/V)}{\partial \vartheta} = \frac{\partial (\Omega_{orig}/V)}{\partial \vartheta} +\sum_{i\in M}n_s^M\frac{\partial m^*_i}{\partial\vartheta},
    \label{eqn of motion}
\end{equation}

 where $\vartheta = \sigma, \zeta,\delta, \omega_0,\rho_0,\phi_0$. The first term corresponds to the original CMF equations of motion, which include the thermal contribution from interacting baryons \cite{Cruz-Camacho:2024odu}, while the second term represents the thermal contribution from the interacting mesons reviewed in this work \cite{Kumar:2025rxj}. Note that non-interacting mesons would not contribute to Eq.~\eqref{eqn of motion}, since their masses would be independent of the mean fields. In the present model, however, the in-medium meson masses depend explicitly on the mean fields and are obtained from the second derivatives of the interaction potential with respect to the corresponding meson fields (see Eq.~\eqref{mass}).

\begin{equation}
    m^{*2}_{ij} = \lim_{\varphi \longrightarrow<\varphi> } -\frac{\partial^2 U}{\partial\phi_i\partial\phi_j},
    \label{mass}
\end{equation}

with $\varphi = \pi,\eta,\eta',K,\omega,\rho,K^*,\phi$, with vacuum expectation value  $<\varphi>=0$.
The resulting effective masses depend explicitly on the scalar and vector mean fields and therefore evolve with both temperature and baryon chemical potential, as illustrated in Fig. \ref{fig:masses}. Note that the effective mass expressions were computed before applying the mean-field approximation.  In the present work, we employ the C4 vector self-interaction scheme; further details can be found in Refs.~\cite{Kumar:2024owe,Kumar:2025rxj}.

\begin{figure}
    \centering
    \includegraphics[
        width=1\linewidth,
        trim={0cm 0cm 0cm 0cm},
        clip
    ]{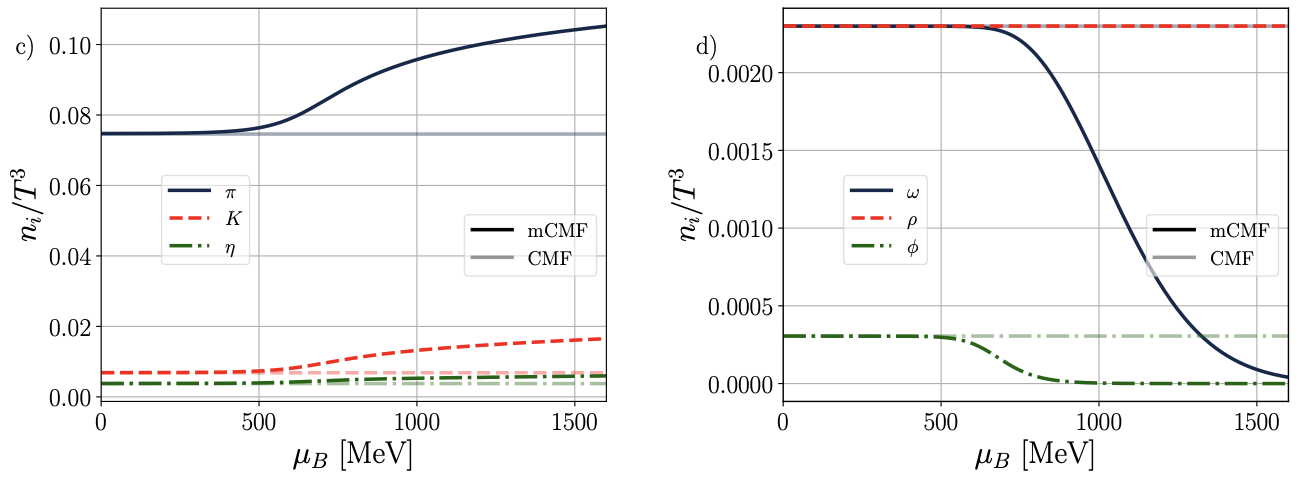}
    \caption{Scaled particle population vs. baryon chemical potential at T = 100 MeV for the non-interacting mesons
(CMF) as well as the medium-modified (interacting) meson (mCMF) cases. Panel c) and d) show, respectively, pseudoscalar and vector mesons. (Figure taken from Ref. \cite{Kumar:2025rxj})}
    \label{fig:population}
\end{figure}

\section*{Result}
\begin{figure}
    \centering
    \includegraphics[width=1\linewidth,
        trim={0cm 0cm 0cm 0cm},
        clip]{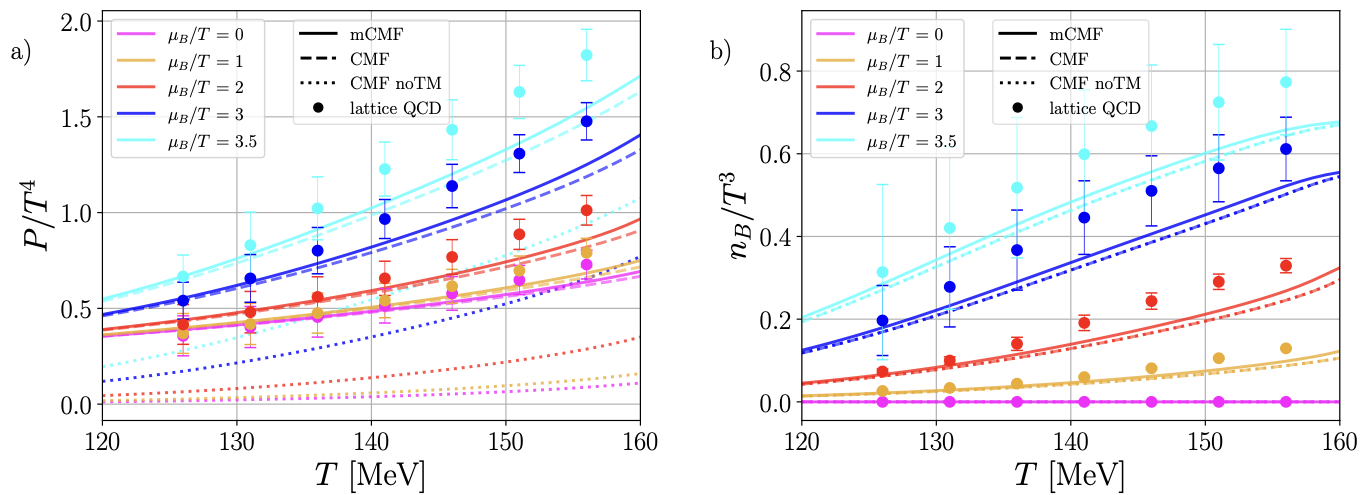}
    \caption{Thermodynamic variables scaled by appropriate powers of temperature vs. temperature for different $\mu_B/T$
ratios for the no meson (CMF noTM), non-interacting mesons (CMF), and interacting mesons (mCMF) cases. Panels are a) scaled pressure, b) scaled (net) baryon number density. We also show state-of-the-art lattice QCD results extrapolated to finite $\mu_B$ \cite{Borsanyi:2021sxv} for comparison.}
    \label{fig:lattice}
\end{figure}

Figure~\ref{fig:masses} shows the in-medium masses for the pseudoscalar mesons ($\pi, K, \eta)$ on panel a), which all exhibit a gradual decrease  as $\mu_B$ increases and show temperature dependence due to the onset of chiral symmetry restoration. In contrast, in panel b), for the vector mesons ($\omega, \phi)$, which show a steep increase with increasing $\mu_B$, indicative of significant modification in dense matter.  Figure~\ref{fig:population} shows the particle population for the interacting pseudoscalar and vector mesons in mCMF compared to the non-interacting ones in CMF results shown as constant lines the new mCMF shows sensitivity with both $\mu_B$ at $T=100$ MeV. Figure~\ref{fig:lattice} compares the thermodynamic quantities of three model variants: CMF without thermal mesons (CMF noTM), CMF with non-interacting thermal mesons (CMF), and the interacting thermal-meson model (mCMF), against lattice-QCD results \cite{Borsanyi:2021sxv}. The panels show a) the scaled pressure, $P/T^4$, and b) the scaled net baryon density, $n_B/T^3$, as functions of temperature for several values of $\mu_B/T$.

\section*{Conclusions}

The new Chiral Mean Field (CMF) model with improved meson description (mCMF) consistently incorporates interacting thermal mesons with in-medium masses. By obtaining the meson masses from the effective chiral Lagrangian before applying the mean-field approximation, thermal mesons contribute directly to the equations of motion through feedback terms, leading to a modified hadronic equation of state. The medium dependence of the meson masses results in noticeable changes in the thermal meson populations, with pseudoscalar and vector mesons exhibiting distinct behavior as functions of baryon chemical potential and temperature. The inclusion of interacting thermal mesons increases the pressure near the crossover region, leading to a significantly better agreement with lattice-QCD results over the considered range of $\mu_B/T$ compared with previous CMF implementations employing non-interacting thermal mesons. These results demonstrate that treating thermal mesons as interacting degrees of freedom provides a more realistic description of hot and dense hadronic matter while preserving the successful features of the CMF framework. Future work will extend the present formalism to include the quark sector and investigate the complete equation of state across the QCD phase diagram for applications to heavy-ion collisions and compact-star mergers.

\section*{Acknowledgments}
This material is based upon work supported by NSF, PCLB, DoE, and NASA

\bibliographystyle{elsarticle-num}
\bibliography{sqm2026_template}



\end{document}